\documentclass[aps,prb,twocolumn,groupedaddress,showpacs,fleqn,floatfix]{revtex4}
\usepackage{amsmath}
\usepackage{epsfig}

\newcommand{\Eq}[1]{Eq.\,(\ref{#1})}
\newcommand{\stress}{\overset{\leftrightarrow}{{\mathbf \sigma}}}
\newcommand{\strain}{\overset{\leftrightarrow}{{\mathbf \epsilon}}}
\begin{document}

\title{Electron-Phonon Interaction in Embedded Semiconductor Nanostructures}

\author{Frank Grosse}
\email{frank.grosse@physik.hu-berlin.de}
\homepage{http://www-semic.physik.hu-berlin.de}
\author{Roland Zimmermann}
\affiliation{Institut f\"ur Physik der Humboldt-Universit\"at
 zu Berlin, Newtonstr. 15, 12489 Berlin, Germany}

\date{\today}

\begin{abstract}
The modification of acoustic phonons in semiconductor
nanostructures embedded in a host crystal is investigated
including corrections due to strain within continuum elasticity
theory. Effective elastic constants are calculated employing {\em
ab initio} density functional theory. For a spherical InAs quantum dot embedded in 
GaAs barrier material, the electron-phonon coupling is calculated. Its strength is 
shown
to be suppressed compared to the assumption of bulk phonons.
\end{abstract}


\maketitle

\section{Introduction}
Semiconductor nanostructures have attracted increasing interest
over the last couple of years. Especially quantum dots (QD) are
proposed as possible candidates in applications like single
photon sources and quantum information devices. Improved growth
techniques nowadays enable the fabrication of small QDs with a
narrow size distribution. Single, isolated QDs embedded
in a semiconductor matrix or in solution are of high interest for
basic research. Since their electronic properties are inevitable
connected to the underlying lattice and its vibrations (phonons)
they offer the possibility to study these fundamental
interactions. While the theory for electron-phonon interaction in bulk
systems is well established, the inhomogeneous nature of
nanostructures leads to strong modifications of the electronic
properties as well as the phonon spectrum. Usually, QDs
are designed such that at least the energetically lowest state is
spatially confined. The existence of boundaries between the
constituting materials and/or to the vacuum introduces a coupling
of the longitudinal and transverse phonon modes even for
isotropic media. Additionally, new types of confined interface
and surfaces modes can occur. Especially the electron-acoustic
phonon interaction is of high interest since it is the source of
so-called pure dephasing of optical excitations in QDs,
as detailed in recent articles~\cite{KAK2005,MZ2004}.

The electron-acoustic phonon interaction is usually treated by
deformation potential coupling, and acoustic phonons are
described as bulk phonons. Only recently some progress has been
made to include phonons in inhomogeneous media in the treatment
of electron-phonon interaction in slab and half space
geometries~\cite{KAK2005}. So far to the best of our knowledge no
investigations exist treating the modified phonons of the
nanostructure itself in the interaction.

The present article provides insights into the role of the phonon
modification due to the nanostructure, including strain.
We restrict ourselves to the simplest case -- a spherical QD --
in order to describe the qualitative effect on the
electron-acoustic phonon coupling. The details of the model and
the corresponding theory are developed in Sec.~\ref{SECMODEL}.
Calculations of strain dependent acoustic phonons on the basis of
{\em ab initio} density functional theory (DFT) calculations are
given in Sec.~{\ref{SECDFT}} which serve as input for the
calculation of the electron-phonon coupling function. Results are
given in Sec.~\ref{SECRESULTS}.

\section{Model and Theory}
\label{SECMODEL} Epitaxial growth methods enable the fabrication
of semiconductor heterostructures with sizes of a few nanometers
only. The different materials are connected pseudomorphically,
i.e. without structural defects. Defects would lead to partial
relaxation of intrinsic strain which results from different
lattice constants of the constituents. The strain itself causes
the change of physical properties compared to relaxed bulk
materials. This includes the electronic (band structure) as well
as the lattice properties (phonons).

The system we study in this article is a spherical shape
inclusion (quantum dot) in an infinite matrix (barrier). This artificial system should
serve as test model for the relevance of non-bulk phonon modes in
the electron-phonon coupling. For the material parameters we
choose the prototype InAs/GaAs system. The first part is devoted
to the description of the acoustic phonons within continuum
elasticity theory followed by the description of the
electron-acoustic phonon coupling.

\subsection{Acoustic phonons within continuum elasticity}
In elastic continuum theory, acoustic phonons follow from the
wave equation of the displacement field ${\bf u}({\bf r})$
\begin{equation}
 -\omega^2 \rho({\bf{r}}) u_j({\bf{r}}) = \sum_k
 \frac{\partial}{\partial x_k} \sigma _{jk} ({\bf{r}}) \,,
\label{EQNPHON}
\end{equation}
here written in Cartesian coordinates $j,k = x, y, z$. Both the
mass density $\rho({\bf r})$ and the stress tensor $\stress({\bf
r})$ are in general spatially dependent. In linear elasticity
theory the stress tensor $\stress$ is proportional to the strain
tensor $\strain$ defined by the spatial derivatives of the
displacement fields, both quantities being related by the fourth order tensor of 
elastic constants $\overset{\leftrightarrow}{C}$ (Hooke's law). For
isotropic systems there are only two independent elastic
constants $C_{11}$ and $C_{44}$, which are related to the well-known
Lam{\'e} coefficients $C_{12} = C_{11} - 2 C_{44} = \lambda$ and
$C_{44} = \mu$. The stress tensor elements are then given by
\begin{equation}
 \sigma_{jk}=\delta_{jk} \left(C_{11} - 2 C_{44} \right)\sum_{l}\frac{\partial
u_{l}}{\partial x_{l}} + C_{44}\left(\frac{\partial
u_{j}}{\partial x_{k}}+ \frac{\partial u_{k}} {\partial
x_{j}}\right).
\end{equation}
The restriction to isotropic media allows to treat the phonons in
large part analytically.

For our spherical model system it is convenient to transform to
spherical coordinates. For the specific electron-phonon coupling
considered below, the only relevant phonon mode has angular momentum zero (breathing 
mode), ${\bf u}({\bf r}) = {\bf e}_r u(r)$, which fulfills the equation
\begin{eqnarray} 
\label{Diff}
0 & = & \rho(r) \omega^2 u(r) - 4 \frac{d C_{44}(r)}{dr}\frac{u(r)}{r} \\
& & + \frac{d}{dr} \left[ C_{11}(r) \left(\frac{du(r)}{dr} +
\frac{2 u(r)}{r} \right) \right] \nonumber
\end{eqnarray}
with elastic constants and mass density having an arbitrary
radial dependence. This one-dimensional second-order differential
equation contains the boundary conditions for a stepwise constant
medium as well, which are
\begin{eqnarray}
u \quad \mbox{and} \quad C_{11}\left(\frac{d u}{dr}\, + \,
\frac{2u}{r}\right) \, - \, 4 C_{44}\frac{u}{r} \quad
\mbox{continuous} \, .
\label{boundary}
\end{eqnarray}
We solve \Eq{Diff} numerically, starting with a regular solution
$u(r) \propto r$ at the origin. The general normalization
condition of the displacement within a large volume $\Omega$
reads \cite{TCG94}
\begin{eqnarray} \label{Norm}
\int_\Omega d^3r \,\rho({\bf r})\, {\bf u}_\nu({\bf r}) \cdot
{\bf u}_{\nu'}({\bf r}) = \delta_{\nu,\nu'} \frac{\hbar}{2
\omega_\nu}\, ,
\end{eqnarray}
where $\nu$ is the discrete mode index. This normalization is
converted into a matching condition at large distances with an
outgoing wave containing a phase shift, as standard in scattering
theory. As a consequence, $\nu$ is replaced by a (continuous)
energy variable $\hbar \omega_\nu \rightarrow E$.

The elastic constants in \Eq{Diff} are not to be taken as the
usual tabulated bulk values for the dot and the barrier material:
The phonon calculation has to start from the reference state
where the crystal is locally under finite strain. Therefore the
spatial dependency is not only due to the change in material but
also due to the spatially varying strain.

\subsection{Effective Elastic Constants from DFT}
\label{SECDFT}

\begin{figure}
 \includegraphics[ width=.45\textwidth]{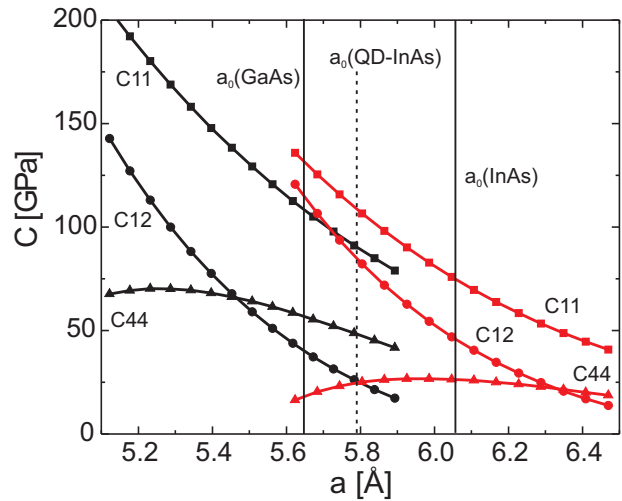}
 \caption{Effective elastic constants for GaAs (black) and InAs (red) in dependence on 
the lattice constant $a$ for hydrostatic deformations calculated by DFT employing 
perturbation theory. The (experimental) equilibrium lattice constants are marked by 
full vertical lines, while the lattice constant of the relaxed InAs QD is given by the 
dashed line.}
 \label{Fig01}
\end{figure}

The usual treatment of continuum elasticity assumes fixed elastic constants. In our 
case, however, the system contains spatial regions where the material is strained. 
Intuitively, one would expect that a material under compression would become harder 
due to the presence of anharmonic terms in the expansion of the free energy density 
around the equilibrium position,
\[
\rho_0 \phi = \frac{1}{2} \sum_{I,J=1}^6 C_{IJ} \epsilon_I \epsilon_J + \frac{1}{3!} 
\sum_{I,J,K=1}^6 C_{IJK}
\epsilon_{I} \epsilon_{J} \epsilon_{K} + \ldots
\]
The equilibrium density is written as $\rho_0$. We restrict ourselves to hydrostatic 
compressions in the following. An extension to general deformations is
straightforward but numerically a far more complex task. Allowing only small 
distorsions of the lattice (an expansion up to second order is sufficient), only the 
bulk modulus $B = 1/3 (C_{11} + 2 C_{12})$ would be a constant quantity. Including up 
to third order in the expansion an analytical derivation gives a linear dependence on 
the strength of the hydrostatic lattice distortion~\cite{BD2005}.

The calculation of elastic constants is possible directly by employing DFT.
Starting from a relaxed unit cell under the
application of finite deformations within density functional
perturbation theory one can calculate the fourth order tensor of
elastic constants. 
The density functional calculations are carried out with the
ABINIT computer code~\cite{GBC+2002,GRV+2005,abinit}. The local density
approximation is applied for the exchange-correlation. Soft norm-conserving 
pseudopotentials are taken from the code of the Fritz-Haber Institute, 
Berlin~\cite{FS99}.
Wave functions are expanded into plane waves with converged
Monkhorst-Pack meshes of $k = 8 \times 8 \times 8$ per $1 \times 1$ unit cell and a 
cutoff energy of $E_{cut} = 36\,\mathrm{Ha} \equiv 979 \,$eV.

The dependence of the calculated elastic constants on the (variable) lattice constant, 
which corresponds to hydrostatic deformation, is displayed in Fig.~\ref{Fig01}. In the 
range from -7\% to +7\% lattice constant variation we find a nearly linear reduction 
of $C_{11}$ and $C_{12}$ with increasing the lattice constant, which is expected from 
a third order expansion~\cite{BD2005}, showing that even for small distorsions the 
assumption of constant elasticity does not hold. The deviation from the linear 
behavior, especially for $C_{44}$, is a clear indication that even higher orders than 
the third one influence significantly the effective elastic constants. Numerical 
deviations from the experimental values are mainly due to the known underestimation of 
the lattice constant in LDA for III-V semiconductors (see Tab.~\ref{Tab01}). Given a 
local static lattice distortion, it is now possible to implement these elastic 
constants in the phonon calculation, which depend on the spatial position due to 
change in material and strain as well.

\subsection{Electron-acoustic phonon coupling}
The electron-phonon interaction has two main contributions:
Deformation potential coupling and piezo-electric coupling. The
second contribution will not be considered here, although some of
the following discussions apply also to this part.

For conduction electrons at the $\Gamma$-point in III-V semiconductors the deformation 
potential couples only to a volume deformation
($\mathrm{div}\, {\bf{u}}$). The interaction Hamiltonian is~\cite{TCG94}
\begin{equation}
H_\mathrm{def}({\bf{r}})=\sum_{\nu} D_C({\bf r}) \left[
b_{\nu}^{\dagger} \mathrm{div}\, {\bf{u}}_{\nu} (
{\bf{r}})+h.c.\right] \label{mainep}
\end{equation}
with the (material-dependent) deformation potential constant for
the conduction band $D_C$ and the phonon creation operator
$b_{\nu}^{\dagger}$. In a QD with confinement functions
$\varphi_n(\mathbf{r})$ of the electron, the relevant
electron-phonon matrix elements are then
\begin{equation} \label{ElPhon}
M^{nm}_\nu = \int d^3 r \, \varphi^*_n(\mathbf{r}) \,
\varphi_m(\mathbf{r})\,D_C(\mathbf{r})  \, {\mathrm{div}}\,
{\bf{u_{\nu}}}({\bf{r}}) \, .
\end{equation}
The spherical symmetry of the QD leads to well-defined angular
quantum numbers of the confinement levels, which select an
appropriate symmetry of the lattice displacement. For electronic
s-s-transitions, only the azimuthal quantum number $l = 0$ is
needed (breathing mode). For the polaronic modification of a
single electronic transition within the independent Boson
model~\cite{Mahan}, the diagonal element in \Eq{ElPhon} is of
central importance. Concentrating on the lowest (s-type)
confinement state $n = 1$, we need to calculate the following
coupling function~\cite{MZ2006}
\begin{equation} \label{Coupling}
f(E)=\sum_{\nu} \delta (E-\hbar \omega_{\nu})\left| M^{11}_\nu
\right|^2\, .
\end{equation}
For instance, the so-called broad band around the zero-phonon
line at $E = 0$ has a shape close to $f(E)/E^3$ at elevated
temperatures (strictly speaking, the hole confinement state
participating in the transition has to be included as well).

\begin{figure}
 \includegraphics[ width=.45\textwidth]{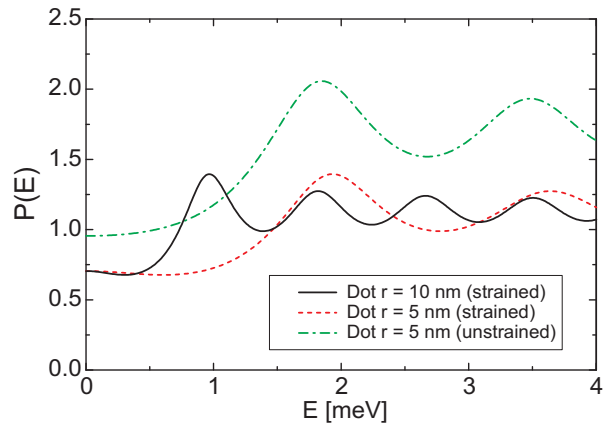}
\caption{Energy dependence of the phonon amplitudes for an InAs
sphere of radius $R$ embedded into GaAs material. The unrealistic
situation without strain is compared with the strained one.}
 \label{Fig02}
\end{figure}

\section{Results and Discussion}
\label{SECRESULTS} In the following we present results for a
spherical QD consisting of pure InAs embedded in infinitely extended GaAs. Under 
hydrostatic compression the coordinates change according to $x_j' = (1-\alpha) x_j$. 
The deformation $\alpha$ can be determined from the static solution ($\omega = 0$) of 
Eq.~\ref{Diff}. Applying the boundary conditions from Eq.~\ref{boundary} accordingly 
leads to 
\begin{eqnarray}
\alpha & = & \alpha_0 \frac{C_{44,\mathrm{GaAs}}}{ C_{44,\mathrm{InAs}}- 
C_{44,\mathrm{GaAs}} + \frac{3}{4} C_{11,\mathrm{InAs}}}.
\end{eqnarray}
with the lattice mismatch $\alpha_0$. For the present
InAs/GaAs model system, it is $\alpha_0 = (a_{0,\mathrm{InAs}} -
a_{0,\mathrm{GaAs}})/a_{0,\mathrm{InAs}} = 6.8$\% using the
experimental  values given in Tab.~\ref{Tab01}. The relaxed
lattice constant $a$ of the dot material (InAs), which is
compressed hydrostatically, is therefore $a_{\mathrm{InAs}} =
a_{0,\mathrm{InAs}} ( 1 - \alpha )$ (last row of
Tab.~\ref{Tab01}, and indicated in Fig.~\ref{Fig01} as dashed
line.)

Outside the dot, the material is compressed in radial
direction but dilated laterally resulting in a zero net volume
change. Since we consider only hydrostatic compression the
elastic properties are taken unchanged for the barrier
material. 

\begin{figure}
 \includegraphics[ width=.45\textwidth]{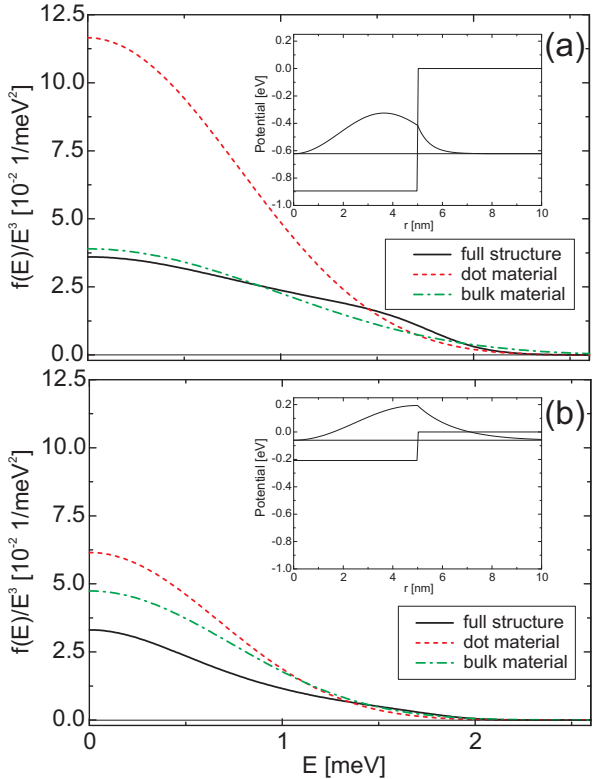}
\caption{Electron-acoustic phonon coupling function. Compared are
(a) the unstrained combination of an InAs QD ($R = 5\,$nm) embedded in GaAs and the 
more realistic (b) strained situation. In each panel the assumption of bulk phonons, 
either
GaAs or InAs, is compared to the full phonon calculation. The
inset shows the electron band offset and the confinement charge
density.} 
\label{Fig03}
\end{figure}

\begin{figure}
 \includegraphics[ width=.45\textwidth]{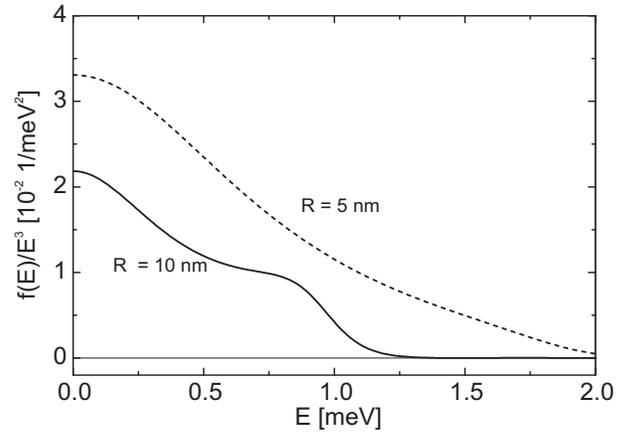}
 \caption{Electron-acoustic phonon coupling function: Compared are two different
radii of the strained InAs QD embedded in GaAs.}
 \label{Fig04}
\end{figure}

We start by looking at the common approximation of using bulk
phonons of the barrier material (index B) throughout. The solution of
\Eq{Diff} gives simply
\begin{equation}
u^{(B)}(r) = j_1 ( k_B r)
\end{equation}
with $j_l(x)$ being the spherical Bessel function of the first
kind, and the wave number is related to the energy via $E = \hbar
v_{lB} k_B$ ($v_{l} = \sqrt{C_{11}/\rho}$ is the longitudinal
sound velocity). Insertion into \Eq{Coupling} leads to
\begin{equation}
f^{(B)}(E) = \frac{E^3 D_{CB}^2}{4 \pi^2 \hbar^3 \rho_B v_{lB}^5}
\left| \int_0^\infty dr \, r^2 \varphi^2_1(r) \, j_0(k_B
r)\right|^2 \, .
\end{equation}
In this approximation, the coupling function (divided by $E^3$)
follows the (squared) Fourier transform of the electron charge
density, which decays on a scale of $E \approx \hbar v_{lB} /R$
($R$ is the QD radius).

For the full problem, the displacement inside the dot (index D) is getting
a prefactor, 
\begin{equation}
u(r) = A(E)\, j_1 (k_D r) \, ,
\end{equation}
which has to be determined from solving Eq.~\ref{Diff} numerically. Although the 
integration in the matrix element extends into the
barrier material as well, it is mainly the variation of $A(E)$
which modifies the coupling function. In order to implement the
change in wave number (sound velocity) as well, we found it
convenient to define a phonon amplitude as
\begin{eqnarray}
P(E) = A(E) \frac{v_{lB}}{v_{lD}} \, .
\end{eqnarray}
By definition, for bulk barrier phonons, $P(E) \equiv 1$ which is
the normalization used in Fig.~\ref{Fig02}.

Whereas for a free standing sphere as well as for a rigidly
clamped sphere, only discrete phonon energies are allowed, the
QD embedded in an {\em infinite} elastic medium allows
for a continuous manifold of phonon energies. However, the
oscillating behavior of the phonon amplitude shown in Fig.~\ref{Fig02} resembles the
eigenmodes of an isolated sphere. A specific feature is the reduction of the phonon 
amplitude at small energies: Including the strain, its overall strength is reduced 
even further. Doubling the size of
the embedded dot from $R = 5$\,nm to $R = 10$\,nm leads to halve
the energetic spacing between maxima, while the overall amplitude
remains the same. This can be understood by noting that within
the applied approximation, neither the static strain field nor
the material density depends on the size of the QD.

\begin{table}
\begin{tabular}{c|ccccccc}
       & $m_e$         & $D_C$       & CBO       &   $\rho$    & $a_0$(Exp) & 
$a_0$(DFT)   \\
       & [$m_0$]       &  [eV]       &  [meV]    & [g/cm$^3$]  &  [\AA] &  [\AA] \\ 
\hline
GaAs   & 0.067         &  -7.17      &  0        &   5.33      &  5.65 &   5.51       
\\
InAs   & 0.026         &  -5.08      &  -894     &   5.66      &  6.06 &   6.05     \\
\hline
s-InAs &  0.026       &  -5.08      &  -208     &   6.50      &  5.79 &  ---     
\end{tabular}
\caption{Parameters used in the calculation for the unstrained 
material~\cite{VMR2001}. Given are also the theoretical lattice constants $a_0$ (DFT). 
Last row: Resulting values for the embedded strained InAs quantum dot (s-InAs).}
\label{Tab01}
\end{table}

The calculation of the electron confinement wave function is done
within the effective mass approximation including mass
discontinuities. All relevant input variables are taken from
existing literature~\cite{VMR2001} and given in Tab.\,\ref{Tab01}. The 
compressive strain in the InAs dot reduces the conduction band offset (CBO) 
significantly. The resulting electron confinement potential is plotted in the 
insets of Fig.~\ref{Fig03} together with the charge density in the lowest 
confinement state, $r^2 \varphi^2_1(r)$. In the strained case 
(Fig.~\ref{Fig03}b), a larger tunneling of the wave function into the barrier 
can be seen. However, this change in the electronic wave function affects the 
electron-phonon coupling only marginal.

The coupling function \Eq{Coupling} of electrons with acoustic
phonons is shown in Fig.~\ref{Fig03} comparing the strain free
case (a) with the improved description including the strain
present in the nanostructure (b). Within each panel the
approximation of bulk phonons using either barrier (GaAs) or dot
(InAs) material parameter are displayed, too. Compared to 'dot-bulk phonons', a strong 
reduction is found, whereas 'barrier-bulk phonons' are much closer to the full 
solution, although missing the existing structure in the coupling function.

The compressive strain results in an increased mass density (see Tab.~\ref{Tab01}), 
increased elastic constants in the dot and to a reduction of the coupling strength 
$\mathrm{div}\, \mathbf{u}(\mathbf{r})$. The shape of the coupling function, however, 
is not much changed when including strain since the maximum positions in the phonon 
amplitude are only slightly shifted. The coupling is reduced further by increasing the 
size of the QD, see Fig.~\ref{Fig04}. Here, the coupling function is
even more structured (sharp drop at around 1 meV) which has its origin in the 
oscillatory variation of the phonon amplitude discussed before.

\section{Conclusions}
Including the details of the phonon characteristics in
semiconductor nanostructures is essential for a quantitative
description of the electron-acoustic phonon coupling. Material
parameters, i.e. mass density and elastic constants, are changed
due to intrinsic strain in pseudomorphically grown
heterostructures and are therefore spatially dependent. Within
continuum elasticity this can be treated by introducing effective
material parameters. There is a continuous energy spectrum for
phonons of nanostructures embedded in an infinite elastic medium.
Compared to the assumption of bulk phonons, an overall reduction
of the electron-acoustic phonon coupling is found due to the
nanostructure as well as due to strain. Resonance-like features
in the phonon modes lead to changes in the line shape of the
energy dependent coupling function which are more pronounced for
mid-size than for small dots.

\begin{acknowledgments}
The work has been done within the framework of SFB 296 of the Deutsche 
Forschungsgemeinschaft. Fruitful discussions with L. Wendler are acknowledged. 
\end{acknowledgments}


\end{document}